# A Driver ASIC for Scientific CCD Detectors Using 180nm Technology

Jie Gao, Dong-xu Yang, Yi Feng, Guang-yu Zhang, Wen-qing Qu, Jian-min Wang, Hong-fei Zhang, Jian Wang, *Senior Member, IEEE*

*Abstract*—In order to achieve the driver function for several types of scientific CCD detector from E2V Co Ltd, and decreasing the size of electronics of CCD detector system, an Application-specified Integrated Circuit (ASIC) was designed. It provides multi-channel clocks and bias voltage for CCD driver. In the ASIC, the clock drivers are made of a clock switch circuit and high voltage amplifier. Two 8-bit current-steering DACs are used to adjust the driver capability and high-level voltage of clocks. The bias drivers are generated by 8-bit current-steering DACs and off-chip operation amplifiers. The Global Foundry 180 nm BCDlite technology is selected to implement this design. The first version of design has been finished and the tests have been done.

## I. INTRODUCTION

Charge-coupled devices (CCD) have been widely used in astronomical telescopes since the 1970s result from their excellent quantum efficiency and low noise etc [1]. A large scale mosaic CCD detector system usually requires a driver with varieties of clocks and biases, while complicated logic control timing is required, which becomes a technical difficulty in the design of CCD detector system.

The CCD detector system must be driven by a massive electronic system, which is often built with discrete components [2][3]. The best way to reduce the size of electronic system is to make the discrete devices of driver circuits to several ASICs [4]. Another benefit is that the ASIC approach can also reduce the total power dissipation for a large scale mosaic CCD detector system. An ASIC has been designed for these purposes. It is used for CCD driver and called BCDA (Bias Clock Driver ASIC), provides multi-channel clocks and bias voltage.

The first version design of BCDA has been finished. In a 4 mm × 4 mm bare chip, six channels of clock circuits, four channels of Bias circuits and some other test circuits are implanted. The ASIC has been manufactured and the tests have been done.

## II. DRIVER ASIC DESIGN

The 180 nm BCDlite technology of Global Foundries is selected to implement the design, due to many CCDs manufactured by E2V Co Ltd required multiple Bias voltage below 30 V and clock signals ranging from 0 V to 16 V. The technology contains both normal 5/6 V CMOS transistors and 10 – 35 V high voltage LDMOS.

The scheme of BIAS circuits is shown in Figure 1. The output current of an 8-bit current-steering DAC generates an voltage drop on an off-chip resistor, $R_1$, which is also connected to the positive terminal of the off-chip high-voltage amplifier, ranging from 1.3 – 6 V. Thus the output voltage of the off-chip amplifier is from 8 – 30 V. The reference circuit of DAC is generated by an on-chip low-voltage amplifier and an off-chip high-precision resistor, $R_0$, so that the reference current is not affected by the technology process.

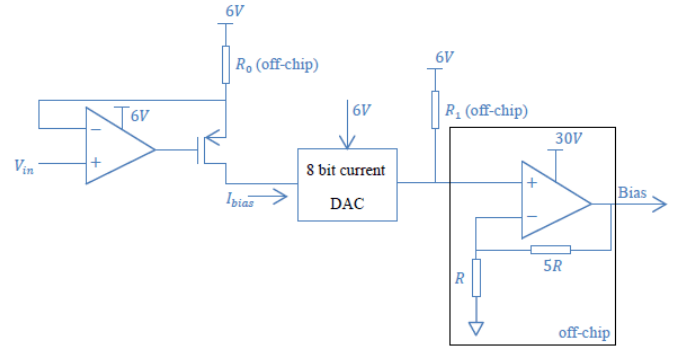

Figure 1 The scheme of BIAS circuits

The scheme of CLOCK circuits is shown in Figure 2. There are two types of CCD clocks, one is parallel clock which is slow and heavily loaded, and the other is serial clock which is fast and light loaded. The designs of two clocks are the same except the output driver capacity. A clock switch with the range of upper's rail voltage from 0V to 16V is designed to generate the clocks. The structure of the voltage generation of upper's rail is same as BIAS circuits shown in Figure 1. The input of the clock switch is single-end LVCMOS signal, which is converted to differential clock signal by an internal single-to-differential circuit. The output current of an 8-bit current-steering DAC is used as tail current of the clock switch to change the driving

This work was supported by the National Natural Science Funds of China under Grant No: 11603023, 11773026, 11728509, the Fundamental Research Funds for the Central Universities (WK2360000003, WK2030040064), the Natural Science Funds of Anhui Province under Grant No: 1508085MA07, the Research Funds of the State Key Laboratory of Particle Detection and Electronics, the CAS Center for Excellence in Particle Physics, the Research Funds of Key Laboratory of Astronomical Optics & Technology, CAS.

The Authors Jie Gao, Dong-xu Yang, Yi Feng, Guang-yu Zhang, Wen-qing Qu, Jian-min Wang, Hong-fei Zhang, Jian Wang are with the University of Science and Technology of China, Jian Wang, State Key Laboratory of Technologies of Particle Detection and Electronics, University of Science and Technology of China, Hefei, Anhui 230026, China (e-mail: Hong-fei Zhang, nghong@ustc.edu.cn; Jian Wang, wangjian@ustc.edu.cn).

capability.

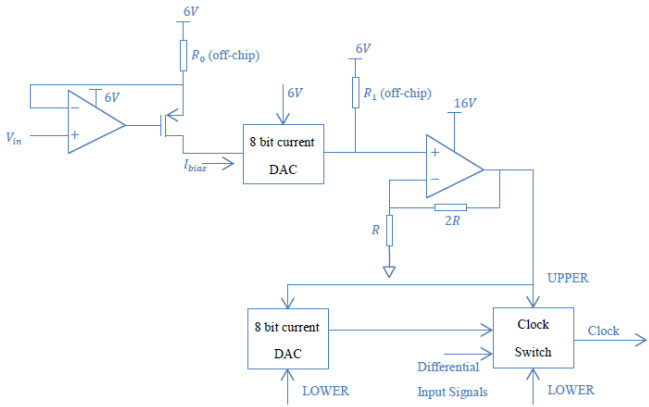

Figure 2 The scheme of CLOCK circuits

The layout of BCDA is shown in Figure 3. The area of a channel of bias, parallel clock, and serial clock is around 690 μm × 183 μm, 784 μm × 344 μm, and 672 μm × 342 μm respectively.

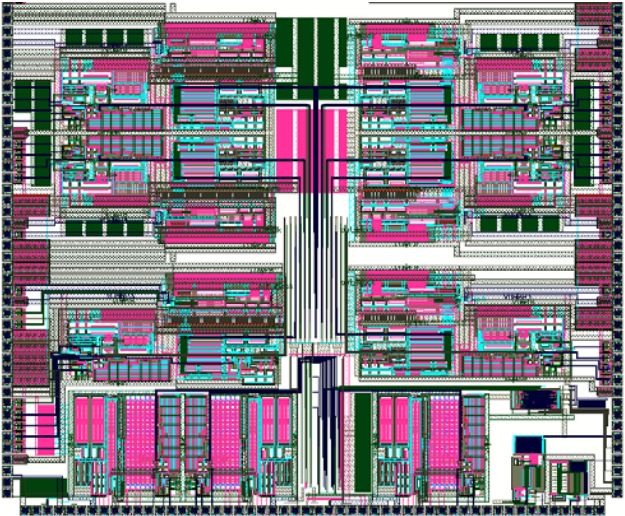

Figure 3 The layout of BCDA

## III. TEST RESULT

Figure 4 shows the results of one of the 10 DACs. All 8-bit DACs for BIAS and CLOCKs work as expected with excellent linearity, stability.

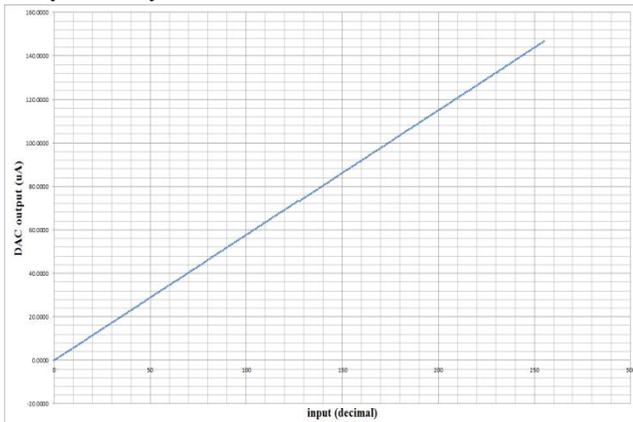

Figure 4 The Results of 8-bit DAC test, Output Current vs. Input Code

The waveform shape of the CLOCKs, including the rise time, the fall time and the switching speed, will directly affect the readout speed and full well capacity of CCD47-20 and CCD230-84. Figure 5 shows the results of CLOCKs with the heaviest load and maximum driving capability.

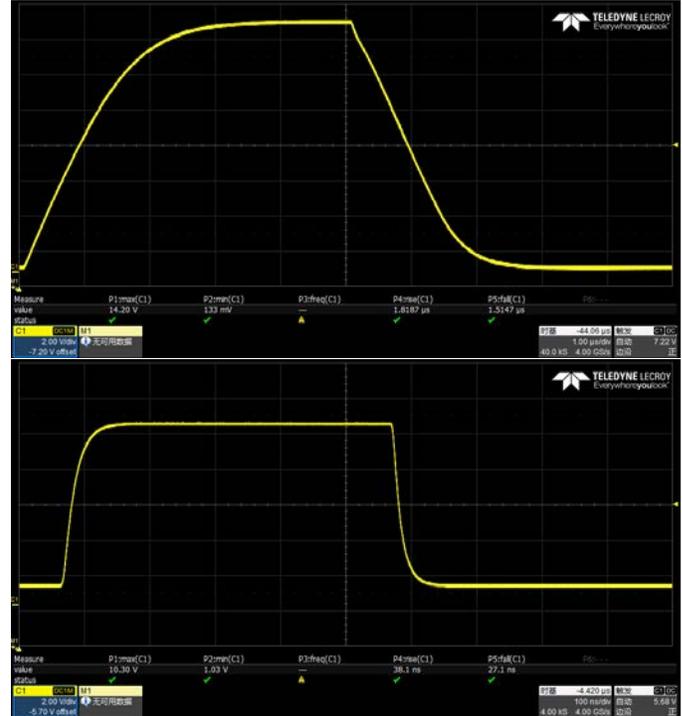

Figure 5 The Results of CLOCKs test, Parallel Clock (upper) and Serial Clock (bottom)

## IV. CONCLUSION

We have implemented the BCDA for CCD driver, which manufactured in 180 nm BCDlite technology of Global Foundries. The BCDA successfully decrease the volume of CCD electronics system, and prove that the ASIC method is a feasible solution for large scale mosaic CCD detector system.